%% file: note1555.tex
\documentclass[11pt]{article}
\usepackage{cite}
\usepackage{epsfig}
\usepackage[hang,center]{subfigure}
\usepackage{lineno,multirow,tabls,amssymb,amsmath}

\newcommand{\BABARPubYear}    {06}

\newcommand{\BABARConfNumber} {010}
\newcommand{\SLACPubNumber} {11984}

\newcommand{\BztoLbppim}        {\ensuremath{\Bz \to \Lbar \proton \pim}\xspace}
\newcommand{\Lambdac}           {\ensuremath{\Lambda_{c}^{+}}\xspace}
\newcommand{\Sigmac}            {\ensuremath{\Sigma_{c}}\xspace}
\newcommand{\Sigmacz}           {\ensuremath{\Sigma_{c}(2455)^{0}}\xspace}
\newcommand{\BzbtoLcpb}         {\ensuremath{\Bzb \to \Lambdac \antiproton}\xspace}
\newcommand{\BmtoLcpbpi}        {\ensuremath{\Bm \to \Lambdac \antiproton \pim}\xspace}
\newcommand{\pKpi}              {\ensuremath{\proton \Km \pip}\xspace}
\newcommand{\Lcp}               {\ensuremath{\Lambdac \antiproton}\xspace}
\newcommand{\splot}             {\ensuremath{_{s}\mathcal{P}lot}\xspace}

\newcommand{\BtoLppi}           {\ensuremath{\Bm \to \Lambdac \antiproton \pim}\xspace}
\newcommand{\LctopKpi}          {\ensuremath{\Lambdac \to \proton \Km \pip}\xspace}
\newcommand{\Lppi}              {\ensuremath{\Lambdac \antiproton \pim}\xspace}
\newcommand{\mLcp}              {\ensuremath{m_{\Lambda_c \proton}}\xspace}
\newcommand{\mLcpi}             {\ensuremath{m_{\Lambda_c \pi}}\xspace}

\newcommand{\msqLcpi}           {\ensuremath{m^{2}_{\Lambda_c \pi}}\xspace}
\newcommand{\msqppi}            {\ensuremath{m^{2}_{\proton \pi}}\xspace}

\input pubboard/babarsym
\def\geant      {\mbox{\tt GEANT4}\xspace} 

\long\def\inst#1{\par\nobreak\kern 4pt\nobreak
    {\it #1}\par\vskip 10pt plus 3pt minus 3pt}

\begin{document}
{\pagestyle{empty}

\begin{flushright}
\babar-CONF-\BABARPubYear/\BABARConfNumber \\
SLAC-PUB-\SLACPubNumber \\
%hep-ex/\LANLNumber \\
July 2006 \\
\end{flushright}

\par\vskip 5cm

\begin{center}
\Large \bf Measurement of the Branching Fractions of the Decays \BzbtoLcpb and \BtoLppi
\end{center}
\bigskip

\begin{center}
\large The \babar\ Collaboration\\
\mbox{ }\\
\today
\end{center}
\bigskip \bigskip

\begin{center}
\large \bf Abstract
\end{center}
We present studies of two-body and three-body charmed baryonic $B$ decays in a sample of 232 million \BB\ pairs collected with the \babar\ detector at the \pep2\ \epem storage ring. The branching fractions of the decays $\bar{B}^{0} \to \Lambda_{c}^{+} \bar{p}$ and $B^{-} \to \Lambda_{c}^{+} \bar{p} \pi^{-}$ are measured to be $(2.15 \pm 0.36 \pm 0.13 \pm 0.56) \times 10^{-5}$ and $(3.53 \pm 0.18 \pm 0.31 \pm 0.92)\times10^{-4}$, respectively. The uncertainties quoted are statistical, systematic, and from the \LctopKpi branching fraction. We observe a baryon-antibaryon threshold enhancement in the $\Lcp$ invariant mass spectrum of the three-body mode and measure the ratio of the branching fractions to be $\BR(\BtoLppi)/\BR(\BzbtoLcpb) = 16.4 \pm 2.9 \pm 1.4$. These results are preliminary.

\vfill
\begin{center}

Submitted to the 33$^{\rm rd}$ International Conference on High-Energy Physics, ICHEP 06,\\
26 July---2 August 2006, Moscow, Russia.

\end{center}

\vspace{1.0cm}
\begin{center}
{\em Stanford Linear Accelerator Center, Stanford University, 
Stanford, CA 94309} \\ \vspace{0.1cm}\hrule\vspace{0.1cm}
Work supported in part by Department of Energy contract DE-AC03-76SF00515.
\end{center}

\newpage
}

\input pubboard/authors_ICHEP2006.tex

\section{\boldmath INTRODUCTION}
\label{sec:Introduction}

\begin{figure}[!b]
\begin{center}
\subfigure[]{\epsfig{file=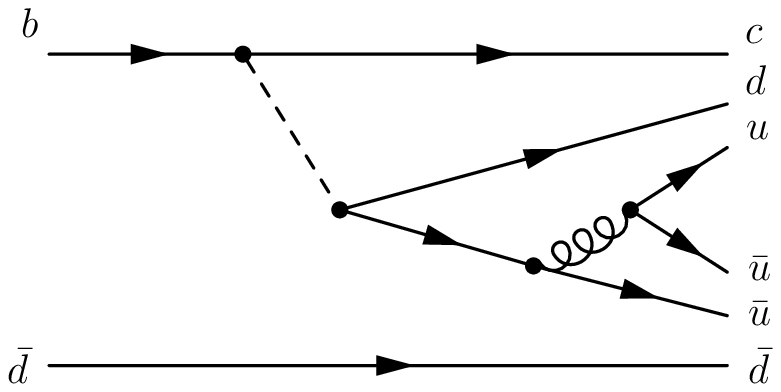,width=0.45\textwidth}}%
\subfigure[]{\epsfig{file=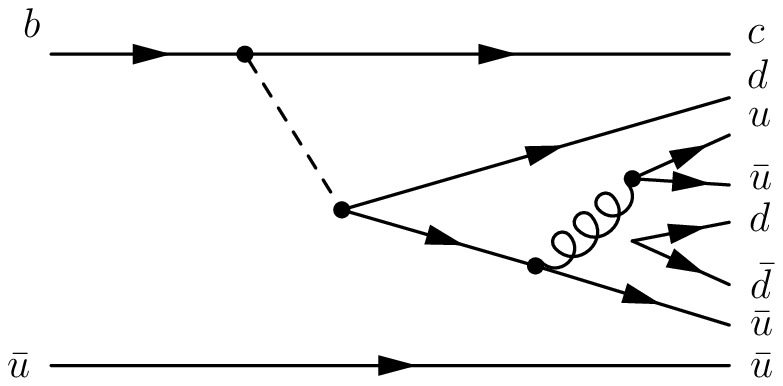,width=0.47\textwidth}}
\renewcommand{\baselinestretch}{1}
\caption{\label{fg:feynmandiag} Feynman diagrams for (a) \BzbtoLcpb and (b) \BtoLppi, in which the \B meson decays weakly via internal \W emission.}
\end{center}
\end{figure}

Charmed baryonic \B decays are experimentally accessible and provide a way to check predictions given by various theoretical models for exclusive baryonic \B decays. There is theoretical interest in the suppression of the two-body baryonic decay rates compared to three-body decay rates and the possible connection to production mechanisms for baryons in \B decays. Analysis of the charmed three-body baryonic \B decay reveals that the invariant mass of the baryon-antibaryon system is peaked near threshold~\cite{ref:belle2004}. Charmless two-body baryonic \B decays (which have not yet been observed~\cite{ref:babarBppb,ref:belleBLamLam}) may be used to measure direct CP violation in the \B system. Their charmed counterparts, however, have branching fractions at least an order of magnitude higher than the charmless modes, and thus can help distinguish between theoretical models that predict the charmless decay rates of \B mesons to baryons. The Feynman diagrams for these decays are shown in Figure~\ref{fg:feynmandiag}, in which the \B meson decays weakly via internal \W emission to $\Lambdac \antiproton (\pi)$.

Charmed baryonic \B decays have recently been measured by the CLEO~\cite{ref:cleo2002} and Belle~\cite{ref:belle2002,ref:belle2003,ref:belle2004} Collaborations. In particular, the Belle Collaboration has measured the branching fractions of the modes\footnote{Throughout this paper, whenever a mode is given, the charge conjugate is also implied.} \BzbtoLcpb (using $85$ million \BB pairs)~\cite{ref:belle2003} and \BmtoLcpbpi (using $152$ million \BB pairs)~\cite{ref:belle2004}: 

\[\BR(\BzbtoLcpb) = (2.19^{+0.56}_{-0.49} \pm 0.32 \pm 0.57) \times 10^{-5}\;\textrm{and}\]
\[\BR(\BtoLppi) = (20.1 \pm 1.5 \pm 2.0 \pm 5.2) \times 10^{-5},\]
where the errors are statistical, systematic, and from the \LctopKpi branching fraction, respectively. \babar\ has collected nearly three times the data used in the Belle analysis of the two-body mode, and we can therefore perform a more precise measurement of this branching fraction. For now, the measurement errors are dominated by the $26\%$ fractional error on $\BR(\LctopKpi) = (5.0 \pm 1.3)\%$~\cite{ref:pdg2004}, but this uncertainty cancels in the ratio of the three-body to two-body branching fractions. 

The excess of events near the baryon-antibaryon production threshold seen by Belle in \BtoLppi has also been observed in \BztoLbppim~\cite{ref:bellecharmless3body} and several $\B \to \proton \antiproton \X$~\cite{ref:belleppbar3body,ref:babarppbarx} modes. In reference~\cite{ref:chengstatus2003} a qualitative explanation of the larger three-body branching fraction in conjunction with this threshold effect is given. In the two-body decay, the invariant mass of the baryon-antibaryon is simply $m_B$, whereas in the three-body decay, the invariant mass of the baryon-antibaryon can be lower, allowing the baryon-antibaryon to form a quasi-resonance near threshold. The third daughter, the meson, carries away much of the energy. The result, regardless of the interpretation of the threshold enhancement, is that the \B favors three-body baryonic decay modes by an order of magnitude over two-body modes.

In this analysis, we measure the branching fractions for \BzbtoLcpb and \BtoLppi and observe the threshold enhancement in the baryon-antibaryon system of the the \BtoLppi mode.

\section{\boldmath THE \babar\ DETECTOR AND DATASET}
\label{sec:babar}
The data used in this analysis were collected with the \babar\ detector at the \pep2\ \epem storage ring. The data sample used comprises an integrated luminosity of $210\invfb$ ($232$ million \BB pairs) collected from \epem collisions at the \FourS resonance. The \babar\ detector is described elsewhere~\cite{ref:babar}. Exclusive \B meson decays are simulated with the Monte Carlo (MC) event generator \evtgen~\cite{ref:EvtGen} and hadronization (e.g. for continuum \qqbar events) is simulated with \jetset74 ~\cite{ref:Jetset}. The detector is modeled using the \geant simulation package~\cite{ref:Geant4}.

\section{\boldmath ANALYSIS METHOD}
\label{sec:Analysis}

\subsection{\boldmath Candidate Selection}
We reconstruct \Lambdac candidates in the decay mode \LctopKpi, applying a geometric constraint on the \pKpi vertex, which is required to have a \chisq probability greater then $0.1\%$. The \pKpi invariant mass must be between $2.275$ and $2.295\gevcc$. The \pKpi candidates are constrained to the mass of the \Lambdac~\cite{ref:pdg2004}, which provides better resolution in the kinematic variable $\DeltaE = E^{*}_B - \sqrt{s}/2$, where $E^{*}_B$ is the \B candidate energy in the \epem center-of-mass (CM) frame and $\sqrt{s}$ is the total CM energy. \Lambdac candidates are then combined in a geometric fit with a \antiproton (and $\pi$) to form a \Bzb (\Bm) candidate for the \BzbtoLcpb (\BtoLppi) mode. The \chisq probability for the fit to the full decay tree must be greater than $0.1\%$. 

Daughter \proton, \kaon, and $\pi$ candidates must be well-reconstructed in the drift chamber and are identified with likelihood-based particle selectors using information from the silicon vertex tracker, drift chamber, and ring-imaging $\check{\textrm{C}}$erenkov detector. Several requirements differ between the two- and three-body modes. The pions in the \BtoLppi mode have lower momenta; therefore, we apply looser drift chamber tracking requirements to improve the efficiency in several areas of the \Lppi Dalitz plane. The daughter particles in both decay modes have very loose particle identification requirements with two exceptions: 1) the pions in \BtoLppi are required to satisfy stronger kaon and electron rejection criteria, and 2) the \B daughter \proton in \BzbtoLcpb must pass a tight constraint on the likelihood that the track is a proton and stronger electron rejection.

We construct a linear (Fisher) discriminant $\mathcal{F}$ from several event-shape variables to provide continuum suppression: $|\cos\theta^{*}|$ ($\theta^{*}$ is the angle of the \B candidate momentum vector with respect to the beam axis in the \epem CM frame), $|\cos\theta^{\B}_{thr}|$ ($\theta^{\B}_{thr}$ is the angle of the \B candidate thrust axis with respect to the beam axis in the \epem CM frame), and the summed momentum of the rest of the charged and neutral particles in the event in nine cones centered around the thrust axis of the \B candidate. The requirement on $\mathcal{F}$ provides powerful background rejection ($72.8\%$) for the two-body mode, but is less effective for the three-body mode ($28.0\%$) due to a larger component of combinatoric \B backgrounds compared to the continuum component. These values were determined by maximizing $N_s/\sqrt{N_s+N_b}$, where $N_s$ is the number of signal events based on signal MC samples and $N_b$ is the number of background events in \DeltaE upper sidebands ($\DeltaE>0.1\gev$ and $5.2<\mes<5.29\gevcc$) in data.

\begin{figure}[htb]
\begin{center}
\subfigure[\Bzb candidates in data for \BzbtoLcpb]{\epsfig{file=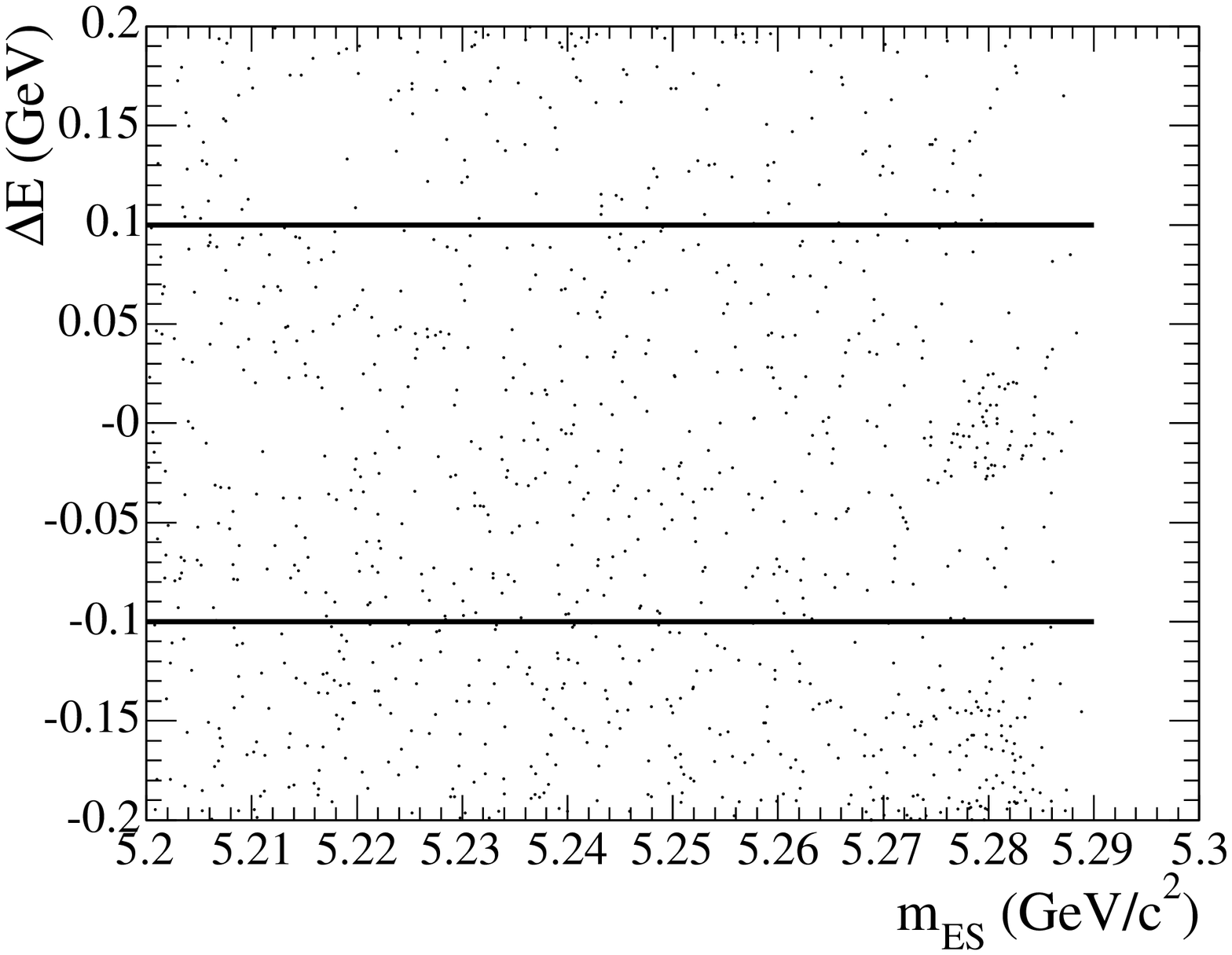,width=0.45\textwidth}}%
\subfigure[\Bm candidates in data for \BtoLppi]{\epsfig{file=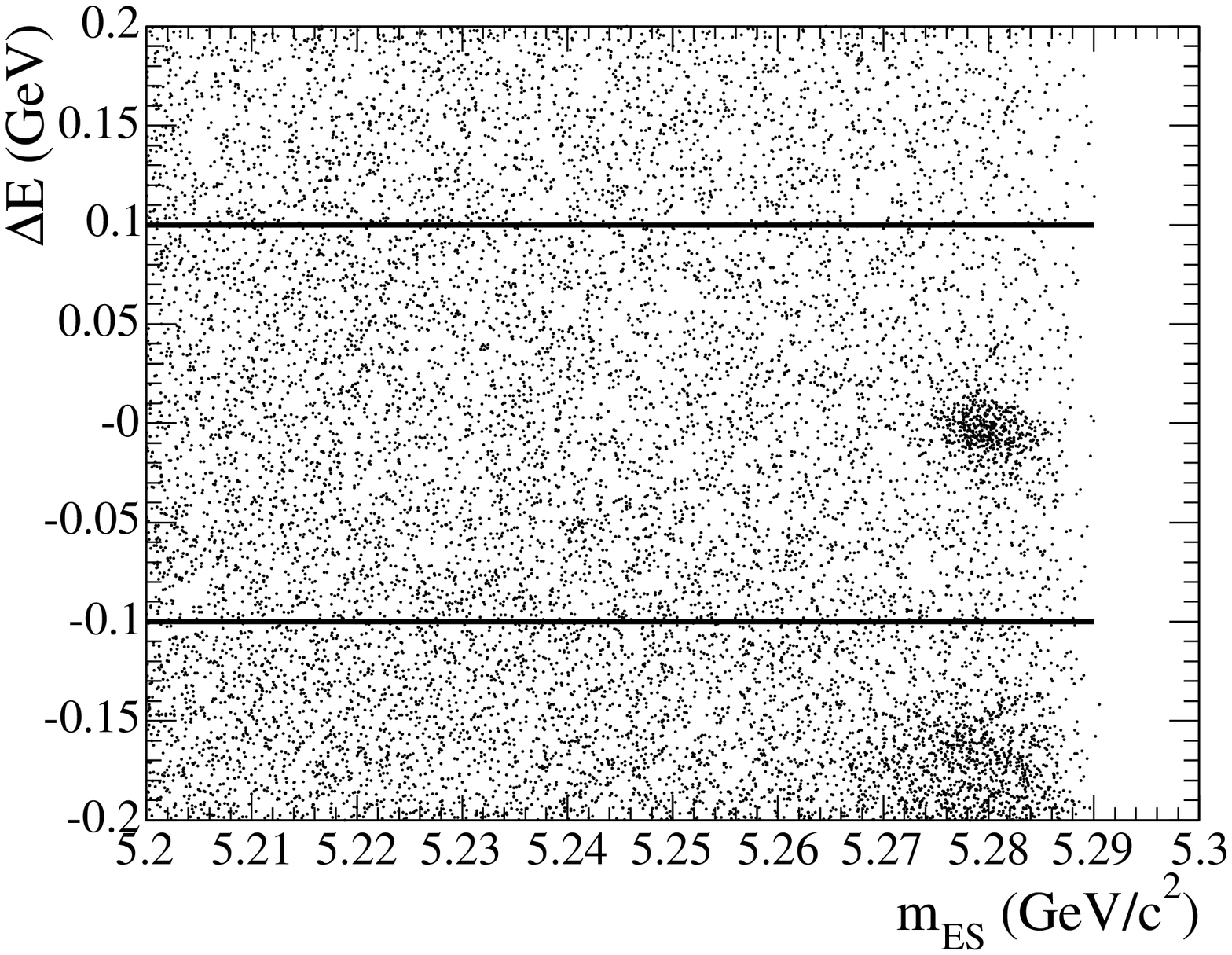,width=0.45\textwidth}}
\renewcommand{\baselinestretch}{1}
\caption{\label{fg:dEvsMes_data} Distribution of \DeltaE vs.\ \mes of \B candidates in data for both the two-body (a) and three-body (b) decay modes. The fit regions are indicated.}
\end{center}
\end{figure}

We identify signal candidates using \DeltaE and the beam-energy-substituted mass \linebreak $\mes=\sqrt{((s/2+\mathbf{p}_i\cdot\mathbf{p}_B)^2/E^2_i-\mathbf{p}^2_B)}$, where $(E_i,\mathbf{p}_i)$ is the four-momentum of the initial \epem system and $\mathbf{p}_B$ is the momentum of the \B candidate, both measured in the laboratory frame. The distribution of \DeltaE vs.\ \mes for both modes is shown in Figure~\ref{fg:dEvsMes_data}. We define the fit region to be $-0.1<\DeltaE<0.1\gev$ and $5.20<\mes<5.29\gevcc$ (also indicated in Figure~\ref{fg:dEvsMes_data}). This excludes the \DeltaE sideband used in the optimization and the region below $-0.1\gev$ in \DeltaE, which contains backgrounds that peak in \mes but are shifted in \DeltaE. These backgrounds are from $\B \to \Lambda_c \proton \pi \pi$ ($\B \to \Lambda_c \proton \pi$) events where a \B daughter $\pi$ is not included in the \B candidate, mimicking the mode of interest: \BtoLppi ($\BzbtoLcpb$). Studies of exclusive MC samples of these backgrounds indicate that much of the contribution is from $\B \to \Sigmac \proton \pi$ ($\B \to \Sigmac \proton$) where a \piz or slow charged $\pi$ from the $\Sigmac \to \Lambda_c \pi$ decay is missed. MC samples comprised of continuum \qqbar events and \B meson decays were studied to rule out any background that peaks in both \DeltaE and \mes.

\subsection{\boldmath \BzbtoLcpb Maximum Likelihood Fit}
In the analysis of the two-body \BzbtoLcpb mode, we did not look at the signal region until the event selection criteria and fit procedures were determined. The efficiency for reconstructing and selecting \BzbtoLcpb candidates is $20.2\%$, and is determined from a fit to the \BzbtoLcpb signal MC sample. A 2-D unbinned maximum likelihood fit is performed in \DeltaE and \mes to extract the number of signal events. The background is described by the product of a linear function in \DeltaE and a threshold function~\cite{ref:argus} in \mes; the signal is described by a single Gaussian distribution in each dimension. All parameters except the \mes threshold are unconstrained in the fit to data. We validate the fitting procedure on a combined sample of signal MC events (over a range of the expected number of signal events) and ``toy'' MC events (generated according to the shape of the continuum and \BB MC background events) to ensure that the fit is robust and unbiased. 

\begin{figure}[tb]
\begin{center}
\epsfig{file=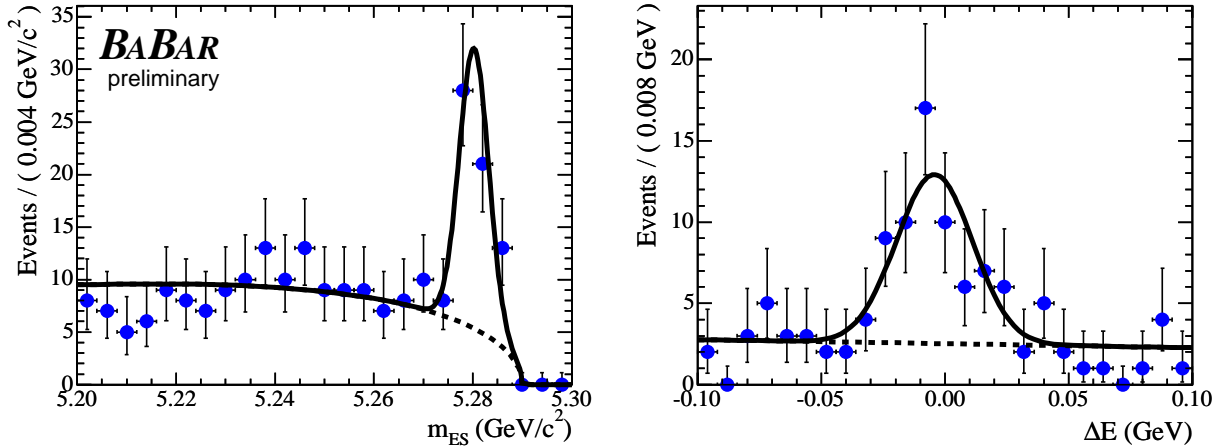,width=\textwidth}%
\renewcommand{\baselinestretch}{1}
\caption{\label{fg:TwoBodyFitResult}Projections of the 2-D fit in \DeltaE and \mes for \Lcp candidates satisfying $|\DeltaE|<0.04\gev$ (left) and $\mes>5.27\gevcc$ (right). The signal yield is $50\pm8$ events, with a significance of $9.4\sigma$.}
\end{center}
\end{figure}

The results of the fit to data are shown in Figure~\ref{fg:TwoBodyFitResult}; we obtain $50\pm8$ signal events and a significance of $\sqrt{-2\ln\left({\cal L}_0/{\cal L}_{max}\right)} = 9.4\sigma$, where ${\cal L}_{max}$ is the maximum likelihood from the fit result and ${\cal L}_0$ is the maximum likelihood when the signal yield is fixed to zero. The mean in \DeltaE is shifted slightly below zero ($-4.2\pm2.7\mev$); this shift is in the appropriate direction given that the \Lambdac mass is constrained to the 2004 PDG value~\cite{ref:pdg2004} which is approximately $1.5\mev$ lower than the most recent measurement~\cite{ref:babarLcmass}. The \DeltaE resolution, $15.4\pm2.1\mev$, is slightly larger than, but consistent with, the resolution in MC ($13.6\pm0.1\mev$).

\subsection{\boldmath \BtoLppi Maximum Likelihood Fit}
\begin{figure}[tb]
\begin{center}
\epsfig{file = 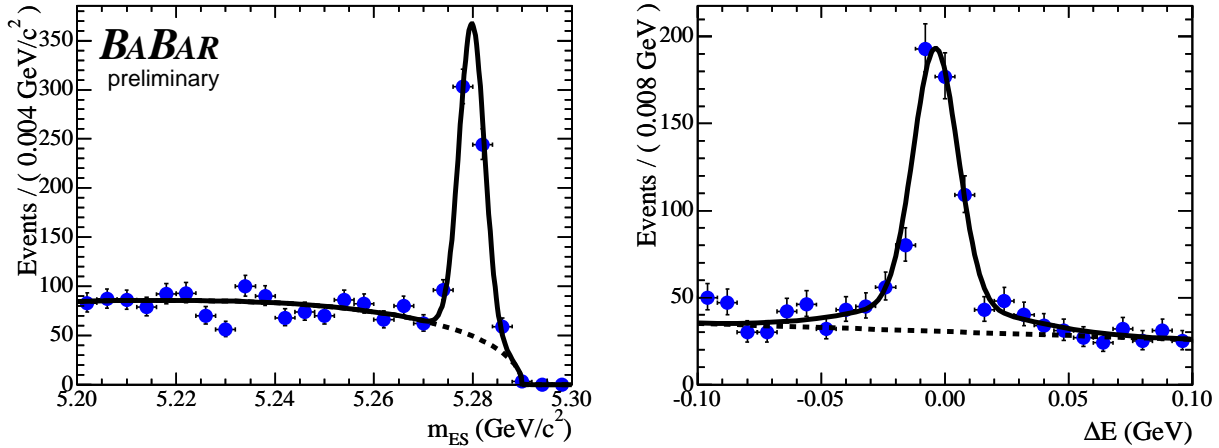, width=\textwidth}
\renewcommand{\baselinestretch}{1}
\caption{\label{fg:ThreeBodyFitResult_Slice} Projections of the 2-D fit in \mes and \DeltaE, for \Lppi candidates satisfying $|\DeltaE|<0.030\gev$ (left) and $\mes>5.27\gevcc$ (right). This 2-D fit is used to extract the likelihood that each event is a signal or background event. The signal yield is $571 \pm 34$ events.}
\end{center}
\end{figure}

For the three-body \BtoLppi mode, a 2-D unbinned maximum likelihood fit is also performed. Again, all parameters except the \mes threshold are unconstrained in the fit to data. The background PDF is the same as in the two-body mode, but the signal PDF consists of a Gaussian in \DeltaE times a Gaussian in \mes, where a correlation is allowed between the two observables. This was not necessary in the two-body mode due to the limited number of signal events. The signal PDF also contains an additional uncorrelated Gaussian component in \DeltaE with the same mean as the correlated Gaussian but an independent width. This signal PDF was chosen from a study of \BtoLppi signal MC events along with extensive studies of various PDFs using a combined sample of signal MC and toy MC events. These studies showed this PDF to have the smallest bias: $-8\pm2$ events for $500$ total signal events (the level of bias is consistent for a range of signal events). The result of the fit to data with this PDF is shown in Figure~\ref{fg:ThreeBodyFitResult_Slice}. The signal yield from the fit is $571 \pm 34$ events and the \DeltaE resolution (RMS) is $19 \pm 3\mev$.

\subsection{\boldmath \BtoLppi Yield Extraction and Efficiency Correction}
We use the \splot technique~\cite{ref:splot} (a sophisticated background subtraction method) to project out the signal and background distributions separately based on the 2-D fit to \DeltaE and \mes. We calculate a signal weight for each event $i$ according to the following equation:
\begin{equation}
W_i = \frac{f_s(\mes_i,\DeltaE_i)+{\mathrm V}_{sb} f_b(\mes_i,\DeltaE_i)}{N_s f_s(\mes_i,\DeltaE_i)+N_b f_b(\mes_i,\DeltaE_i)},
\end{equation}
where $W_i$ is the \splot weight, $N_s (N_b)$ is the number of fitted signal (background) events, and $f_s (f_b)$ is the signal (background) PDF. ${\mathrm V}_{sb}$ is the off-diagonal element of a $2\times2$ covariance matrix calculated directly from data, with all parameters fixed to their fitted values except for the signal and background yields. A background weight for each event can be calculated in an analogous manner. The result of this method is that each event is assigned a signal and background weight, which can be plotted for any quantity that is uncorrelated with \DeltaE and \mes. The quantities of interest that satisfy this requirement are the invariant masses $m_{d_id_j}$, where $d_i$ is any of the \B daughters \Lambdac, \antiproton, $\pim$. The correlations of \DeltaE and \mes with these quantities are less than $5\%$. The \splot method relies on using the events in the entire fit region to provide good sampling of both signal and background. However, (background) events that have an invariant \Lppi mass far from the mass of the \B meson have a different kinematically allowed Dalitz region than (signal) events with an invariant \Lppi mass close to $m_\B$. We calculate $m_{d_id_j}$ with a \B mass constraint so that all of the \B candidates in the fit region lie in the same Dalitz region.

\begin{figure}[tb]
\begin{center}
\epsfig{file = 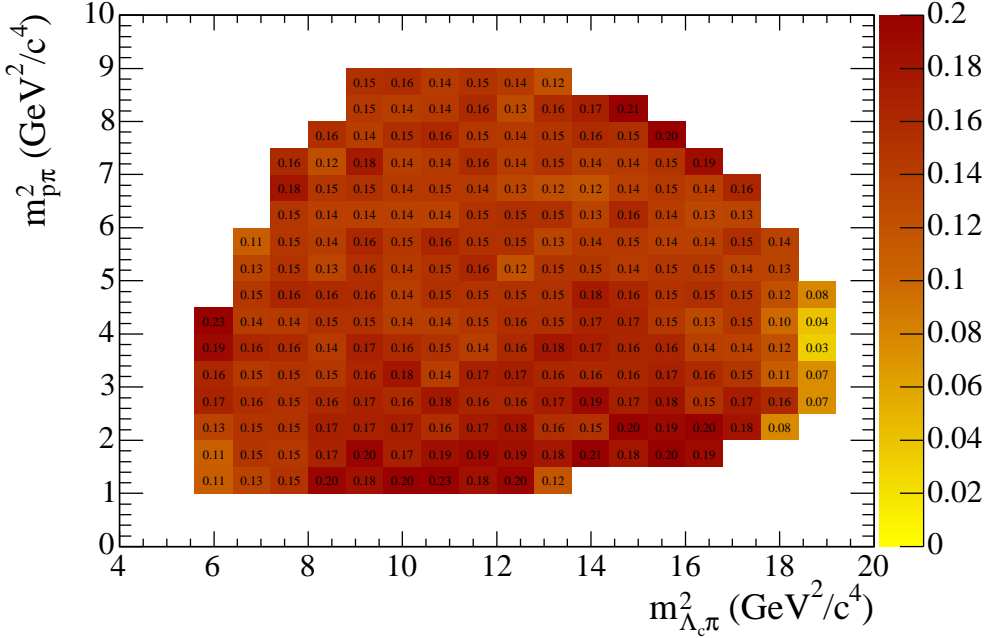, width=0.8\textwidth}
\renewcommand{\baselinestretch}{1}
\caption{Binned efficiency for \msqppi vs. \msqLcpi in the kinematically allowed region of the Dalitz plane.}
\label{fg:DalitzEfficiency}
\end{center}
\end{figure}

The detection efficiency for \BmtoLcpbpi events varies significantly across the Dalitz plane. Therefore, using the average nonresonant MC efficiency ($15.3\%$) to calculate the branching fraction for this mode is insufficient. Instead, an efficiency correction is applied to each signal event based on its location in the Dalitz plane. We divide the physical region into 215 equal-size bins and determine the efficiency in each bin; a plot of this efficiency for \msqppi vs. \msqLcpi is shown in Figure~\ref{fg:DalitzEfficiency}. There are noticeable deficiencies in the lower left (right) corners of the \Lppi Dalitz plane, where the $\pi$ ($\Lambda_c$) candidates have low momentum in the \B rest frame. The looser tracking requirements on the pions help to compensate for this effect, but do not eliminate it entirely. We build on the \splot formalism to individually correct each event by an additional weight, $1/\varepsilon_{\alpha\beta}$, where $\varepsilon_{\alpha\beta}$ is the efficiency in bin $(\alpha,\beta)$. We define an ``effective'' efficiency ($\varepsilon_{\textrm{eff}}$) as the signal yield from the fit divided by the number of \splot-weighted, efficiency-corrected events. The effective efficiency for selecting \B candidates in the three-body mode is $\varepsilon_{\textrm{eff}} = 14.2\%$.

\section{\boldmath SYSTEMATIC STUDIES}
\label{sec:Systematics}

Various sources of systematic uncertainties have been investigated, including those related to the total number of \BB pairs in data, the method used to determine the efficiency from MC, and the fitting procedures. These are summarized for both modes in Table~\ref{tb:systematics}. Note that for the branching fraction measurement of \BzbtoLcpb, the statistical error dominates over the total systematic uncertainty.

\begin{table}[tb]
\caption{Summary of the contributions to the total systematic uncertainty. The total is determined by adding the uncertainty from each source in quadrature.}
\begin{center}
\begin{tabular}{c | c | c}
\multirow{2}{*}{Source} & \multicolumn{2}{c}{Systematic Uncertainty} \\ \cline{2-3}
 & \BzbtoLcpb & \BtoLppi \\ \hline \hline
$N_{\BB}$ & $1.1\%$ & $1.1\%$ \\
MC statistics & $1.0\%$ & $4.7\%$\\
Dalitz binning & -- & $2.0\%$\\
Tracking & $3.0\%$ & $5.2\%$\\
PID & $4.7\%$ & $1.2\%$\\
Fitting & $2.2\%$ & $4.5\%$\\ \hline
Total & $6.0\%$ & $8.7\%$ \\
\end{tabular}
\end{center}
\label{tb:systematics}
\end{table}

The systematic uncertainty on the number of \BB\ pairs produced by \babar\ is $1.1\%$.

There are several sources of systematic uncertainty related to the efficiency determinations. The statistical uncertainty due to the number of signal MC events contributes a $1.0\%$ systematic error on the efficiency. Tracking efficiency systematic errors are based on studies of $\tau$ decays, which yield an uncertainty of $0.8\%$ per track. However, this is reduced to $0.6\%$ for the higher momentum \B daughter \proton in the two-body mode and increased to $1.4\%$ for the lower momentum pions in the three-body mode. Particle identification is determined using large control samples, which may differ from the modes we are investigating due to the higher multiplicities of these charmed baryonic \B decays and other subtleties. Differences between the momentum spectra and angular distributions of the daughter particles compared to those in the control samples are used to assess a systematic uncertainty on the efficiency due to particle identification. In \BzbtoLcpb, we assign a $2.5\%$ systematic uncertainty to the \B daughter \antiproton, and a $(1.5 \:\textrm{to}\: 1.7)\%$ uncertainty on the other daughter particles. In \BtoLppi, the systematic uncertainty due to particle identification varies from $(0.1 \:\textrm{to}\: 0.9)\%$ per track; the total is $1.2\%$.

The fitting systematics are studied by varying the background shape in \DeltaE and the endpoint of the threshold function in \mes. This yields a systematic uncertainty of $2.2\%$ for the two-body mode and $0.9\%$ for the three-body mode. The branching fraction measurement of \BtoLppi has additional systematic uncertainties due to the signal PDF. We assign $4.3\%$ due to the fit bias, the source of which is mostly in the the tails of \DeltaE. This systematic uncertainty compensates for the inability of the MC to accurately simulate the behavior of the events in these tails. For the three-body mode, we perform the fit to data with and without a correlation between \DeltaE and \mes in the signal PDF, yielding a systematic uncertainty of $1.1\%$.

\section{\boldmath RESULTS}
\label{sec:Physics}

The branching fraction of \BzbtoLcpb~\cite{ref:moriond06}, measured with the sample of 232 million \BB\ pairs, is
\[\BR(\BzbtoLcpb) = (2.15 \pm 0.36 \pm 0.13 \pm 0.56) \times 10^{-5},\]
where the errors are statistical, systematic, and from the \LctopKpi branching fraction, respectively. The significance of the signal is $9.4\sigma$. This measurement is consistent with a previous Belle measurement of $(2.19^{+0.56}_{-0.49} \pm 0.32 \pm 0.57) \times 10^{-5}$ made with $85$ million \BB\ pairs. The systematic uncertainty is much lower ($6\%$ compared to $15\%$) than that for the Belle measurement. We also find this measurement to be consistent with the predicted limit from reference~\cite{ref:chengpolemodel2003}: \BR(\BzbtoLcpb) $\lesssim 7.9 \times 10^{-6}\,|g / 5|^2$, where $|g|=6\textrm{--}10$.

We calculate the total branching fraction of \BmtoLcpbpi as follows:

\begin{equation}
\begin{split}
\BR(\BmtoLcpbpi)_{tot} &= \frac{(1+b)\sum\limits_i \dfrac{W_i}{\varepsilon_i}}{N_{\BB}\times\BR(\LctopKpi)}\\
                       &= (3.53 \pm 0.18 \pm 0.31 \pm 0.92)\times10^{-4},
\end{split}
\end{equation}
where the fit bias $b$ is $1.6\%$, $W_i$ is the signal \splot weight and $\varepsilon_i$ is the efficiency for event $i$, and $N_{\BB}$ is the number of \BB pairs. The uncertainties are statistical, systematic, and the error on the \LctopKpi branching fraction, respectively. This measurement is $3.5\,\sigma$ higher (assuming Gaussian statistics) than the Belle measurement of $\BR(\BmtoLcpbpi)_{tot} = (2.01 \pm 0.15 \pm 0.20 \pm 0.52)\times10^{-4}$. An examination of the Dalitz plot shows a systematic trend in that we measure consistently larger branching fractions in all regions.

\begin{figure}[tbp]
\begin{center}
\epsfig{file = 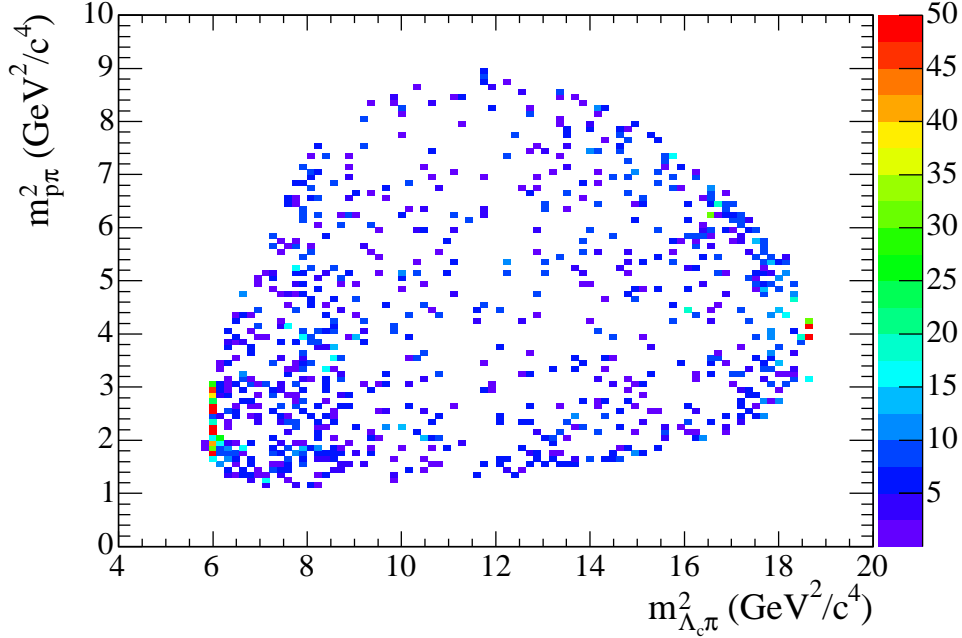, width=0.8\textwidth}
\renewcommand{\baselinestretch}{1}
\caption{\msqppi vs. \msqLcpi for signal \B candidates with \splot and efficiency correction weights applied. Bins with negative population are suppressed.}
\label{fg:sWeights_dalitz}
\end{center}
\end{figure}

\begin{figure}[tbp]
\begin{center}
\epsfig{file = 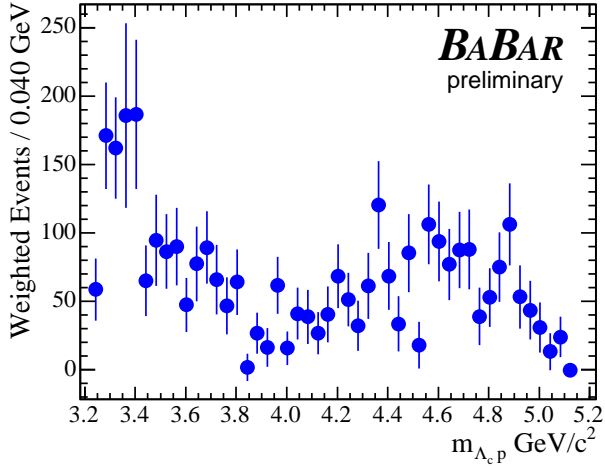, width=0.5\textwidth}
\renewcommand{\baselinestretch}{1}
\caption{\label{fg:sWeights_mLcp} Dalitz plot projection onto the \mLcp axis with the requirement $\mLcpi>2.6\gevcc$, removing the contribution from the \Sigmacz. \splot weighted, efficiency-corrected signal events are shown. The baryon-antibaryon threshold enhancement is visible near $3.3\gevcc$.}
\end{center}
\end{figure}

The \Lppi Dalitz plane in data is shown in Figure~\ref{fg:sWeights_dalitz} with \splot weights and efficiency corrections applied to each \B candidate. We project this onto the \mLcp axis with the requirement $\mLcpi>2.6\gevcc$ (to remove the contribution from the \Sigmacz) in Figure~\ref{fg:sWeights_mLcp}. We observe a baryon-antibaryon threshold enhancement in the \Lcp invariant mass spectrum, confirming the large body of evidence supporting the existence of these threshold enhancements in three-body baryonic \B decays.

We also report the ratio of the branching fractions of \BtoLppi to \BzbtoLcpb:
\begin{equation}
\frac{\BR(\BtoLppi)}{\BR(\BzbtoLcpb)} = 16.4 \pm 2.9 \pm 1.4.
\end{equation}
The systematic uncertainties on the number of \BB pairs, the \Lambdac daughter \proton and \kaon tracking, and the \Lambdac daughter \kaon and \B daughter \proton particle identification all cancel, as does the uncertainty on $\BR(\LctopKpi)$. This ratio is consistent with theoretical predictions.

\section{\boldmath SUMMARY}
\label{sec:Summary}

We report the branching fractions of two charmed baryonic \B decay modes. Table~\ref{tb:bfs} compares the yields, efficiencies, and branching fractions of the two modes. The total three-body branching fraction measured is significantly larger than that measured by Belle, but is still consistent with (and perhaps provides stronger evidence for) the observation that the three-body mode is enhanced over the two-body mode. The measurement of the ratio of three-body to two-body branching fractions and the observation of the baryon-antibaryon threshold enhancement aid in theoretical interpretations of baryon production in \B decays.

\begin{table}[tb]
\caption{Comparison of the yields, efficiencies (effective for the three-body decay), and branching fractions for \BzbtoLcpb and \BtoLppi.}
\begin{center}
\begin{tabular}{c | c | c | c}
Mode & Signal yield & $\varepsilon_{(\textrm{eff})}$ & \BR\\ \hline \hline
\BzbtoLcpb & $50\pm8$ & $20.2\%$ & $(2.15 \pm 0.36 \pm 0.13 \pm 0.56) \times 10^{-5}$ \\
\BtoLppi & $571 \pm 34$ & $14.2\%$ & $(3.53 \pm 0.18 \pm 0.31 \pm 0.92)\times10^{-4}$ \\
\end{tabular}
\end{center}
\label{tb:bfs}
\end{table}

\section{ACKNOWLEDGMENTS}
\label{sec:Acknowledgments}

\input pubboard/acknowledgements

\end{document}

%% file: pubboard/authors_ICHEP2006.tex
\begin{center}
\small

The \babar\ Collaboration,
\bigskip

%% author list as of 01-Jul-2006 (596 authors)
%
{B.~Aubert,}
{R.~Barate,}
{M.~Bona,}
{D.~Boutigny,}
{F.~Couderc,}
{Y.~Karyotakis,}
{J.~P.~Lees,}
{V.~Poireau,}
{V.~Tisserand,}
{A.~Zghiche}
\inst{Laboratoire de Physique des Particules, IN2P3/CNRS et Universit\'e de Savoie,
 F-74941 Annecy-Le-Vieux, France }
% ...

\end{center}\newpage

%% file: pubboard/acknowledgements.tex
We are grateful for the 
extraordinary contributions of our \pep2\ colleagues in
achieving the excellent luminosity and machine conditions
that have made this work possible.
The success of this project also relies critically on the 
expertise and dedication of the computing organizations that 
support \babar.
The collaborating institutions wish to thank 
SLAC for its support and the kind hospitality extended to them. 
This work is supported by the
US Department of Energy
and National Science Foundation, the
Natural Sciences and Engineering Research Council (Canada),
Institute of High Energy Physics (China), the
Commissariat \`a l'Energie Atomique and
Institut National de Physique Nucl\'eaire et de Physique des Particules
(France), the
Bundesministerium f\"ur Bildung und Forschung and
Deutsche Forschungsgemeinschaft
(Germany), the
Istituto Nazionale di Fisica Nucleare (Italy),
the Foundation for Fundamental Research on Matter (The Netherlands),
the Research Council of Norway, the
Ministry of Science and Technology of the Russian Federation, 
Ministerio de Educaci\'on y Ciencia (Spain), and the
Particle Physics and Astronomy Research Council (United Kingdom). 
Individuals have received support from 
the Marie-Curie IEF program (European Union) and
the A. P. Sloan Foundation.